\documentclass[5p,twocolumn,times,sort&compress]{elsarticle}

\usepackage{lineno,hyperref}
\usepackage{graphicx}
\usepackage{dcolumn}
\usepackage{bm}
\usepackage{rotating}
\usepackage{longtable}
\usepackage{float}
\usepackage{eucal}
\usepackage{xcolor}

\modulolinenumbers[5]

\begin{document}

\begin{frontmatter}







\title{\centering {$R$-matrix analysis of $^{22}$Ne($\alpha$, n)$^{25}$Mg and $^{22}$Ne($\alpha$, $\gamma$)$^{26}$Mg reaction}}

\author[label1]{Rajkumar Santra}

\address[label1]{Department of Nuclear and Atomic Physics, Tata Institute of Fundamental Research, Mumbai-400005, India.}

\begin{abstract}
The $^{22}$Ne($\alpha$, n)$^{25}$Mg and its competing channel $^{22}$Ne($\alpha$, $\gamma$)$^{26}$Mg has an major influence on neutron flux in weak s-process nucleosynthesis path in low mass AGB stars and massive stars of mass (M$\geq$ 10M$_\odot$). So the ratio rate of this two competing reaction control the neutron flux in weak s-process nucleosynthesis. Various experiment has been performed to study the properties of nuclear states of $^{26}$Mg to evaluate rate of $^{22}$Ne+$\alpha$ reaction rate and corresponding rate from these studies vary by up to a factor of 500 in the astrophysical relevant temperature. The recent evaluation by Philip et al. of $^{22}$Ne($\alpha$, n)$^{25}$Mg reaction rate using most recent nuclear data of $^{26}$Mg from number of sources shows similar result with previous estimation for $^{22}$Ne($\alpha$, $\gamma$)$^{26}$Mg but got lower rate for $^{22}$Ne($\alpha$, n)$^{25}$Mg reaction due to updated nuclear data. Also Philip et al. suggested that rate based on full $R$-matrix modeling will required to take into accounted the interference effects between distant levels and sub-threshold resonance. In present work full $R$-matrix calculation has been performed for $^{22}$Ne($\alpha$, n)$^{25}$Mg and $^{22}$Ne($\alpha$, $\gamma$)$^{26}$Mg reaction based on fitting the $^{22}$Ne($\alpha$, n)$^{25}$Mg reaction data of Jaeger et al; in energy range 0.8 to 1.45 MeV and updated nuclear data of $^{26}$Mg states. The $R$-matrix fitting for the $^{22}$Ne($\alpha$, n)$^{25}$Mg reaction nicely explain experimental data in 0.8 to 1.45 MeV energy range by changing spin, parity of E$_x$ = 11.784 and 11.63 MeV states from 1$^-$ to 0$^+$.

\end{abstract}


\date{\today}
\end{frontmatter}
Almost half of the elements heavier than iron are synthesis via slow (over time scales of thousands of years) neutron capture reaction on stable isotope in s-process nucleosynthesis path \cite{Pignatari}. 
The s-process manly activated in Asymptotic Giant Branch (AGB) stars, seeded by $^{56}$Fe iron. The $^{13}$C($\alpha$, n)$^{16}$O and $^{22}$Ne($\alpha$, n)$^{25}$Mg reaction are main source of neutrons for s-process. 
For low mass AGB stars with $1M_\odot \leq M \leq 4M_\odot$ both this two reaction activated He inter-shell during the Thermal Pulses contribute as a smaller neutron source and responsible for synthesis elements of Atomic masses A $\approx$ 90 - 209. But for massive stars 
($M \geq 8M_\odot$) only $^{22}$Ne($\alpha$, n)$^{25}$Mg reaction is main source of neutron which synthesis isotopes of mass A $\approx$ 60 - 90 is so call weak s-process nucleosynthesis. Also +ev Q-value reaction 
$^{22}$Ne($\alpha$, $\gamma$)$^{26}$Mg is act as a competing reaction that decide neutron release from $^{22}$Ne($\alpha$, n)$^{25}$Mg reaction. So constraining the rate of this two reaction has high sensitive for
weak s-process nucleosynthesis.

So many direct \cite{Jaeger} and \cite{Jayatissa,Ota, Giesen} indirect experimental measurement has been performed to extracted the spin, parity, partial widths ($\Gamma_\alpha, \Gamma_n, \Gamma_\gamma$) to constraining the rate of $^{22}$Ne($\alpha$, n)$^{25}$Mg
and $^{22}$Ne($\alpha$, $\gamma$)$^{26}$Mg reactions. Recently rate of this two reaction has been re-evaluate \cite{Philip} based upon updated nuclear data from a number of sources and they suggested that an $R$-matrix modeling will required 
due to lack of uncertainty of spin, parity of so many relevant states of $^{26}$Mg as well as interference effects same partials waves between two states.

In this context an multilevel R-matrix analysis has been performed on available cross-section data of $^{22}$Ne($\alpha$, n)$^{25}$Mg reaction including interference effect in energy range E$_\alpha^{c.m}$ $\approx$ 0.8 to 1.45 MeV range 
and extrapolate up to 0.57 MeV energy. Main aim is to constraining spin-parity and study interference effect of states of $^{26}$Mg in excitation energy  E$_x$ $\approx$ 11.319 to 11.828 MeV energy range.

\section{R-matrix analysis of $^{22}$Ne($\alpha$, n)$^{25}$Mg reaction}
The behavior of excitation function for $^{22}$Ne($\alpha$, n)$^{25}$Mg reaction was mainly determined by the resonance capture process through the several resonance states of compound nucleus $^{26}$Mg. In this work an $R$-matrix calculation has been performed to describe capture data and constrain the spin, parity and width of the states of $^{26}$Mg. 

\begin{figure}
\begin{center}
\includegraphics[height=8.0cm, width=9.0cm]{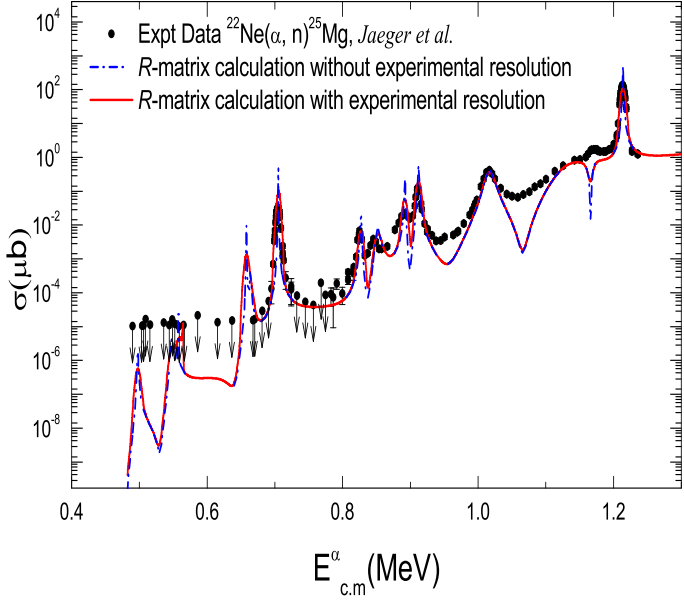}
\caption{\label{fig1} $R$-matrix calculation with literature reported resonance parameters. Filled symbols represent the direct measurement data taken from Ref. \cite{Jaeger}}
\end{center}
\end{figure}


\begin{table}
\centering
\caption{Summary of resonance parameters used in $R$-matrix calculation with literature reported values \cite{Philip} for comparison with direct measurement data of $^{22}$Ne($\alpha$, n)$^{25}$Mg reaction.}
\label{tab2}
\begin{tabular}{ccccccccccccc}     \hline \hline
E$_x$& E$_r$&J$^\pi$&$\Gamma_\alpha$(eV)&$\Gamma_n$(keV) \\ 
(MeV)&(MeV)&Literature \cite{Philip}&Literature\cite{Philip}&Literature\cite{Philip} \\ \hline
11.828&1.214&2$^+$&0.18&1.1 \\ 
11.7847&1.169&1$^-$&8.0 $\times$ 10$^{-3}$&24.5 \\ 
11.749&1.1456& 1$^-$ &0.02&64 \\ 
11.63&1.016&1$^-$ &2.4$\times$10$^{-4}$&13.5 \\ 
11.526&0.9112&1$^-$&4.3$\times$10$^{-4}$&1.8 \\ 
11.508&0.894&1$^-$&1.2$\times$10$^{-4}$&1.27 \\ 
11.461&0.847&3$^-$&7.9$\times$10$^{-6}$&6.55 \\
11.441&0.827&3$^-$ &5.5$\times$10$^{-6}$ &1.47 \\ 
11.3196&0.7056&1$^-$ &5.5$\times$10$^{-6}$&0.132 \\
11.272&0.65&3$^-$ &9.2$\times$10$^{-8}$&1.81 \\
11.258&0.644&2$^+$ &1.0$\times$10$^{-6}$&0.41 \\
11.171&0.557&2$^+$ &1.9$\times$10$^{-8}$&0.03 \\ 
11.169&0.552&3$^-$ &4.4$\times$10$^{-10}$&1.94 \\
11.163&0.549&2$^+$ &2.7$\times$10$^{-9}$&5.31 \\
11.112&0.498&2$^+$ &4.3$\times$10$^{-10}$&2.095 \\ 
11.084&0.470&2$^+$ &5.7$\times$10$^{-11}$&- \\ 
10.9491&0.3351&1$^-$ &3.0$\times$10$^{-14}$&30 \\ \hline
\end{tabular}
\end{table}

$R$-matrix modeling of $^{22}$Ne($\alpha$, n)$^{25}$Mg reaction has been performed using AZURE2 code \cite{Azuma}. This code was developed based on the theory developed of Lane and Thomas \cite{Lane} and of Vogt \cite{Vogt}. 
In $R$-matrix formalism, radial space is divided into two disting regions --an internal region extended up to a radius R$_c \approx r_0(A_p^{1/3} + A_t^{1/3})$ and an external region above R$_c$. 
A choice of radius for entrance and exit channels is needed for the model calculation. R$_c$ of the two channels have been obtained through $\chi^2$ minimization. 
However, as channel radius is not a free parameter in the model, we performed a grid search on the channel radius by changing the value in small steps and varying 
the parameters to get the fit. The chosen channel radii values are 5.37 $fm$ for the $^{22}$Ne + $\alpha$ channel and 4.21 $fm$ for the $^{25}$Mg + n channel. 
During calculation in AZURE2, the energy resolution of the system also accounted.

\begin{table}
\centering
\caption{Summary of resonance state parameters of $^{26}$Mg used for $R$-matrix calculation obtained from Litarature and $R$-matrix fit of $^{22}$Ne($\alpha$, n)$^{25}$Mg data with E$_\alpha$=0.8 to 1.3 MeV.}
\label{tab3}
\begin{tabular}{cccccccccccc}     \hline \hline
E$_x$& E$_r$&J$^\pi$&$\Gamma_\alpha$(eV)&$\Gamma_n$(keV) \\ 
(MeV)&(MeV)& & &  \\ \hline
10.9491&0.335&3$^-$ &3.0$\times$10$^{-14}$\cite{Jayatissa}&0.125\cite{} \\ 
11.084&0.470&2$^+$ &5.7$\times$10$^{-11}$\cite{Jayatissa}& 0.175\cite{}  \\
11.112&0.498&2$^+$ &4.3$\times$10$^{-10}$\cite{Ota1} &2.095 \cite{Massimi}  \\
11.163&0.549&2$^+$ &2.7$\times$10$^{-9}$\cite{Ota1} &5.31\cite{Massimi}  \\
11.169&0.552&3$^-$ &4.4$\times$10$^{-10}$\cite{Ota1} &1.94\cite{Massimi} \\
11.171&0.557&2$^+$ &1.9$\times$10$^{-8}$\cite{Ota1} &0.01\cite{Massimi} \\ 
11.272&0.65&2$^+$ &1.0$\times$10$^{-6}$\cite{Philip} &0.410\cite{Massimi} \\
11.278&0.71&3$^-$ &9.2$\times$10$^{-8}$\cite{Philip} &1.81\cite{Massimi} \\
11.319&0.705&1$^-$ & {\bf 5.1429$\times$10$^{-5}$} & {\bf 0.452}  \\
11.4401&0.827&3$^-$ &{\bf 4.59$\times$10$^{-6}$} & {\bf 0.700} \\
11.458&0.847&3$^-$& {\bf 9.0014$\times$10$^{-6}$} &{\bf 15.588} \\
11.506&0.894&1$^-$& {\bf 1.399$\times$10$^{-4}$} & {\bf 15.346} \\
11.525&0.911&1$^-$&{\bf 2.529$\times$10$^{-4}$}& {\bf 0.5209} \\ 
11.63&1.016& \bf{0$^+$ }& {\bf 9.106$\times$10$^{-3}$} & {\bf 14.341}\\ 
11.759&1.145& 1$^-$ &{\bf 0.0502}& {\bf 139.935} \\ 
11.784&1.169&\bf{0$^+$} &{\bf 23.952 $\times$ 10$^{-3}$}&{\bf 17.226} \\ 
11.8276&1.214&2$^+$ &{\bf 0.210657}& { \bf 1.144} \\ \hline  
\end{tabular}
\end{table}


\begin{table}
\centering
\caption{Summary of estimated ANC from literature reported spectroscopic factors of $^{26}$Mg bound states}
\label{tab1}
\begin{tabular}{ccccccccccccc}     \hline \hline
E$_x$& l&J$^\pi$&S& c$_b$ & C \\ 
(MeV)& & & & fm$^{-1/2}$&fm$^{-1/2}$  \\ \hline
g.s &0&$0^+$ &0.418\cite{Ota}&4.57$\times$10$^{+4}$&2.95$\times$10$^{+4}$  \\
1.81&2&2$^+$ &0.0915\cite{Anantaraman}&2.35$\times$10$^{+4}$&7.10$\times$10$^{+3}$ \\
2.94&2&2$^+$&0.0211\cite{Anantaraman} &2.16$\times$10$^{+4}$&3.13$\times$10$^{+3}$ \\
3.59&0&0$^+$&0.291\cite{Anantaraman} &3.61$\times$10$^{+4}$&1.94$\times$10$^{+4}$ \\
4.32&4&4$^+$&0.046\cite{Anantaraman} &3.16$\times$10$^{+4}$&6.78$\times$10$^{+2}$ \\
4.33&2&2$^+$&0.12\cite{Ota}&2.03$\times$10$^{+4}$&7.032$\times$10$^{+3}$ \\
4.84&2&2$^+$&0.11\cite{Ota}&2.02$\times$10$^{+4}$ &6.69$\times$10$^{+3}$ \\ 
4.9&4&4$^+$&0.2\cite{Ota} &3.07$\times$10$^{+4}$ &1.37$\times$10$^{+3}$ \\ 
4.97&0&0$^+$&0.281\cite{Anantaraman}&3.62$\times$10$^{+4}$ &1.91$\times$10$^{+4}$ \\ 
5.47&4&4$^+$&0.05\cite{Ota}&3.04$\times$10$^{+4}$ &6.79$\times$10$^{+2}$ \\ 
5.71&4&4$^+$&0.04\cite{Ota} &9.80$\times$10$^{+4}$&1.96$\times$10$^{+3}$ \\
6.745&2&2$^+$&0.22\cite{Ota}&4.17$\times$10$^{+4}$ &1.95$\times$10$^{+4}$ \\  
6.876&3&3$^-$&0.17\cite{Ota}&2.81$\times$10$^{+4}$ &1.15$\times$10$^{+4}$ \\  
7.348&3&3$^-$&0.5\cite{Ota}&3.14$\times$10$^{+4}$ &2.22$\times$10$^{+4}$ \\  
8.036&2&2$^+$&0.2\cite{Ota}&6.93$\times$10$^{+4}$ &3.09$\times$10$^{+4}$ \\  
8.937&2&2$^+$&0.14\cite{Ota}&1.91$\times$10$^{+5}$ &7.14$\times$10$^{+4}$ \\  
9.325&2&2$^+$&0.38\cite{Ota}&4.73$\times$10$^{+5}$ &2.91$\times$10$^{+5}$ \\  
9.371&4&4$^+$&0.38\cite{Ota}&9.2$\times$10$^{+4}$ &5.67$\times$10$^{+4}$ \\  
9.856&2&2$^+$&0.1\cite{Ota}&6.81$\times$10$^{+6}$ &2.153$\times$10$^{+6}$\\  
10.573&1&1$^-$&1& 1.92$\times$10$^{+33}$&1.92$\times$10$^{+33}$ \\ \hline  
\end{tabular}
\end{table}

\begin{table*}
\centering
\caption{Summary of bound and resonance state parameters of $^{26}$Mg used for $R$-matrix calculation obtained from Literature and $R$-matrix fit of Jaeger et. al., data $^{22}$Ne($\alpha$, n)$^{25}$Mg data with E$_\alpha$=0.8 to 1.3 MeV.}
\label{tab7}
\begin{tabular}{ccccccccccccc}     \hline \hline
E$_x$& E$_r$&J$^\pi$&$\Gamma_\alpha$(eV)/ANC(fm$^{-1/2}$)&$\Gamma_n$(keV)/ANC(fm$^{-1/2}$) & $\Gamma^{E^f}_\gamma$(eV) \\ 
(MeV)&(MeV)& & &  \\ \hline
g.s &-&$2^+$ &2.95$\times$10$^{+4}$&- \\
1.81&-&2$^+$ &7.10$\times$10$^{+3}$&- \\
2.94&0.470&2$^+$ &3.13$\times$10$^{+3}$&- \\
3.59&0.470&0$^+$ &1.94$\times$10$^{+4}$&- \\
4.32&0.470&4$^+$ &6.78$\times$10$^{+2}$&- \\
4.33&-&2$^+$ &7.032$\times$10$^{+3}$&- \\
4.84&-&2$^+$ &6.69$\times$10$^{+3}$&- \\ 
4.9&-&4$^+$ &1.37$\times$10$^{+3}$&- \\ 
4.97&-&0$^+$ &1.91$\times$10$^{+4}$&- \\ 
5.47&-&4$^+$ &6.79$\times$10$^{+2}$&- \\ 
5.71&-&4$^+$ &1.96$\times$10$^{+3}$&- \\
6.745&-&2$^+$ &1.95$\times$10$^{+4}$&- \\  
6.876&-&3$^-$ &1.15$\times$10$^{+4}$&- \\  
7.348&-&3$^-$ &2.22$\times$10$^{+4}$&- \\  
8.036&-&2$^+$ &3.09$\times$10$^{+4}$&- \\  
8.937&-&2$^+$ &7.14$\times$10$^{+4}$&- \\  
9.325&-&2$^+$ &2.91$\times$10$^{+5}$&- \\  
9.371&-&4$^+$ &5.67$\times$10$^{+4}$&- \\  
9.856&-&2$^+$ &2.153$\times$10$^{+6}$&- \\  \hline 
10.573&-&1$^-$ &1.92$\times$10$^{+33}$&-&0.094$^{g.s}$ \cite{Deboer} \\  
& & &  &  & 0.106$^{4.97}$\cite{Deboer} \\\hline  
10.696&-&4$^+$ &3.5$\times$10$^{-46}$ \cite{LONGLAND} &-&3$^{1.81}$ \cite{LONGLAND} \\  
10.717&-&1$^-$ &2.53$\times$10$^{-36}$\cite{Talwar}&-&3(15)(g.s) \cite{Philip} \\  \hline 
10.805&-&1$^-$ &1.11$\times$10$^{-22}$ \cite{Lotay}&-&0.16$^{g.s}$\cite{Deboer} \\  
& & & & & 0.56$^{1.81}$\cite{Deboer} \\\hline   
10.819&-&0$^+$ &3.18$\times$10$^{-21}$ \cite{Lotay}&-&3$^{g.s}$\cite{Philip} \\  
10.943&-&2$^+$ &5.0$\times$10$^{-11}$(UL)&30 \cite{MASSIMI}&6.5$^{\bf{g.s}}$ \cite{MASSIMI} \\\hline   
10.9491&0.3351&1$^-$ &3.0$\times$10$^{-14}$\cite{Jayatissa}&0.125\cite{} & 0.26$^{g.s}$\cite{Deboer} \\ 
 & & & & & 1.07$^{1.81}$ \cite{Deboer}  \\ 
 & & & & & 0.25$^{2.93}$ \cite{Deboer}  \\ 
& & & & & 0.09$^{3.58}$ \cite{Deboer}  \\ 
& & & & &0.20$^{4.33}$ \cite{Deboer}  \\ \hline 
11.084&0.470&2$^+$ &5.7$\times$10$^{-11}$ \cite{Jayatissa}& 0.175\cite{} & 3$^{4.31}$ \cite{Philip} \\
11.112&0.498&2$^+$ &4.3$\times$10$^{-10}$\cite{Ota1} &2.095 \cite{Massimi}& 1.7$^{1.81}$\cite{Massimi}  \\
11.163&0.549&2$^+$ &2.7$\times$10$^{-9}$\cite{Ota1} &5.31\cite{Massimi}& 2.8\cite{Massimi} \\
11.169&0.552&3$^-$ &4.4$\times$10$^{-10}$\cite{Ota1} &1.94\cite{Massimi}& 3.3$^{1.81}$\cite{Massimi} \\
11.171&0.557&2$^+$ &1.9$\times$10$^{-8}$\cite{Ota1} &0.01\cite{Massimi}& 5 $^{5.47}$\cite{Massimi}\\ 
11.272&0.65&2$^+$ &1.0$\times$10$^{-6}$\cite{Philip} &0.410\cite{Massimi}&2.2$^{0.0}$\cite{Massimi} \\
11.278&0.71&3$^-$ &9.2$\times$10$^{-8}$\cite{Philip} &1.81\cite{Massimi}&0.3$^{1.81}$\cite{Massimi}) \\
11.301&0.71&2$^+$ &1.53$\times$10$^{-5}$\cite{Philip} &-&3$^{1.81}$\cite{Philip} \\
\bf{11.319}&0.7056&1$^-$ & {\bf 5.1429$\times$10$^{-5}$} & {\bf 0.452} &3$^{g.s}$\cite{Philip} \\
11.3277&0.7056&1$^-$ &1.8$\times$10$^{-6}$\cite{Philip} &-&2.2$^{g.s}$\cite{Massimi} \\
11.343&0.7056&2$^+$ &1.1$\times$10$^{-6}$\cite{Philip} &-&1$^{g.s}$\cite{Massimi}  \\
\bf{11.4401}&0.827&3$^-$ &{\bf 4.59$\times$10$^{-6}$} &{\bf 0.700}&3$^{1.81}$\cite{Philip} \\
\bf{11.458}&0.847&3$^-$&\bf{9.0014$\times$10$^{-6}$}&\bf{15.588}&3$^{1.81}$\cite{Philip} \\
11.5002&0.894&1$^-$&1.95$\times$10$^{-1}$\cite{Philip}&-&3$^{g.s}$\cite{Philip} \\
\bf{11.506}&0.894&1$^-$&\bf{1.399$\times$10$^{-4}$}&\bf{15.346}&3$^{g.s}$\cite{Philip} \\
\bf{11.525}&0.9112&1$^-$&\bf{2.529$\times$10$^{-4}$}&\bf{0.5209}&3$^{g.s}$\cite{Philip} \\ 
\bf{11.63}&1.016& \bf{0$^+$ }&\bf{9.106$\times$10$^{-3}$}&\bf{14.341}&3$^{g.s}$ \cite{Philip}\\ 
\bf{11.759}&1.1456& 1$^-$ &\bf{0.0502}&\bf{139.935}&3$^{g.s}$ \cite{Philip}\\ 
\bf{11.784}&1.169&\bf{0$^+$} &\bf{23.952 $\times$ 10$^{-3}$}&\bf{17.226}&3$^{1.81}$\cite{Philip} \\ 
\bf{11.8276}&1.214&2$^+$ &\bf{0.210657}&\bf{1.144}&3$^{1.81}$\cite{Philip} \\ \hline  
\end{tabular}
\end{table*}

In this present work, initially R-matrix calculation has been performed $^{22}$Ne($\alpha$, n)$^{25}$Mg using the recently updated spin, parity and partial widths of resonance states (E$_x$=10.9 to 12.82 MeV) reported in Ref. \cite{Philip} for the observed resonances
in $^{22}$Ne($\alpha$, n)$^{25}$Mg reaction for E$_{\alpha}^{c.m}$= 0.57 to 1.45 MeV range and the resonance states parameters of $^{26}$Mg used in calculation are listed in Table \ref{tab2} and results of $R$-matrix calculation (without and with experimental resolution correction) for cross-section of $^{22}$Ne($\alpha$, n)$^{25}$Mg reaction compare with experimental data of Ref \cite{Jaeger} as shown in Fig. \ref{fig1}. With including experimental resolution correction in $R$-matrix calculation still data are not well reproduced.   

\begin{figure}
\begin{center}
\includegraphics[height=8.0cm, width=9.0cm]{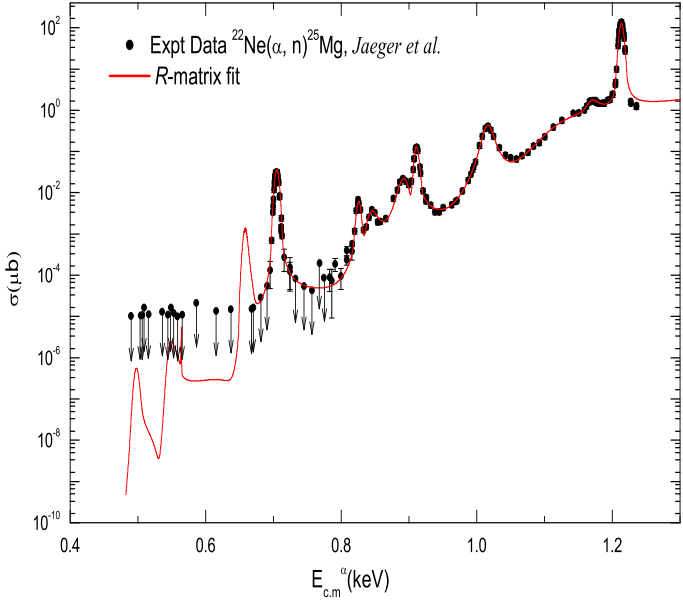}
\caption{\label{fig2} (Color Online) $R$-matrix fit with changing spin and parity shown in Table 2. Filled symbols represent the direct measurement data taken from Ref. \cite{Jaeger} }
\end{center}
\end{figure}

Now to better explain experimental data, we adjust resonance energy and widths of populated all the states with excitation energy range 11.319 to 12.828 MeV (E$_{\alpha}^{c.m}$=0.7 to 1.45 MeV) and by adjusting this parameters we reproduced the data form 0.7 to 0.98 MeV energy region but fails to reproduced data 0.98 to 1.2 MeV energy region due to strong destructive interference between same J$^\pi$ states. Now spin, parity of E$_x$=11.63 and 11.784 MeV resonances are changing form 1$^-$ to 0$^+$ and $\Gamma_\alpha$, $\Gamma_n$ left as free parameters for all the states with excitation energy range 11.319 to 12.828 MeV. The resultant $R$-matrix fit nicely explain the experimental data 
in 0.8 to 1.45 MeV energy region as shown in Fig. \ref{fig2}. The fitted parameters are listed in Table \ref{tab7}. 

\begin{figure}
\begin{center}
\includegraphics[height=6.0cm, width=8.5cm]{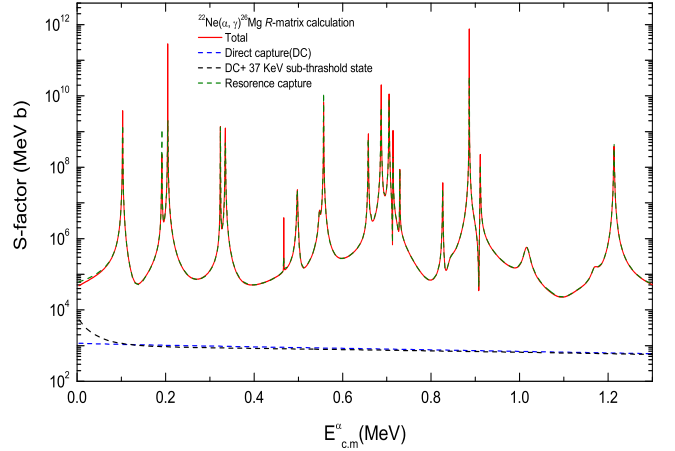}
\caption{\label{fig3} (Color Online) $R$-matrix calculation for $^{22}$Ne($\alpha$, $\gamma$)$^{26}$Mg capture reaction.}
\end{center}
\end{figure}

\section{Estimation of ANC of bound states $^{26}$Mg for direct capture calculation of $^{22}$Ne($\alpha$, $\gamma$)$^{26}$Mg reaction }
The Asymptotic Normalization Coefﬁcients(ANC) define the amplitude of the tail of bound state wave-function for two-body configurations. In low energy radiative capture reaction, capture occurs via the tail of bound state wave function and the ANC determined magnitude of direct capture cross-sections\cite{}. It is related to the
spectroscopic factor of the two-body conﬁguration as

\begin{equation}
C= \sqrt{S} \times c_b
\end{equation}

Where S is the spectroscopic factor of two body configurations. c$_b$ is single particle ANC. The spectroscopic factors of different bound states of $^{26}$Mg are taken from literature \cite{Ota, Anantaraman} which was determined from $^{22}$Ne($^{6}$Li, d)$^{26}$Mg transfer reactions. The c$_b$ are calculated using the code FRESCO (ver. 3.2) \cite{Thompson} code. In FRESCO calculation $\alpha$ binding potentials parameters are use same vale for which spectroscopic factors reported in \cite{Ota, Anantaraman}. The resultant $\alpha$-single particle ANC and total ANC for different bound states of $^{26}$Mg are listed in Table \ref{tab1}.

\section{R-matrix calculation of $^{22}$Ne($\alpha$, n)$^{25}$Mg and $^{22}$Ne($\alpha$, $\gamma$)$^{25}$Mg reaction}
\begin{figure}
\begin{center}
\includegraphics[height=6.0cm, width=8.5cm]{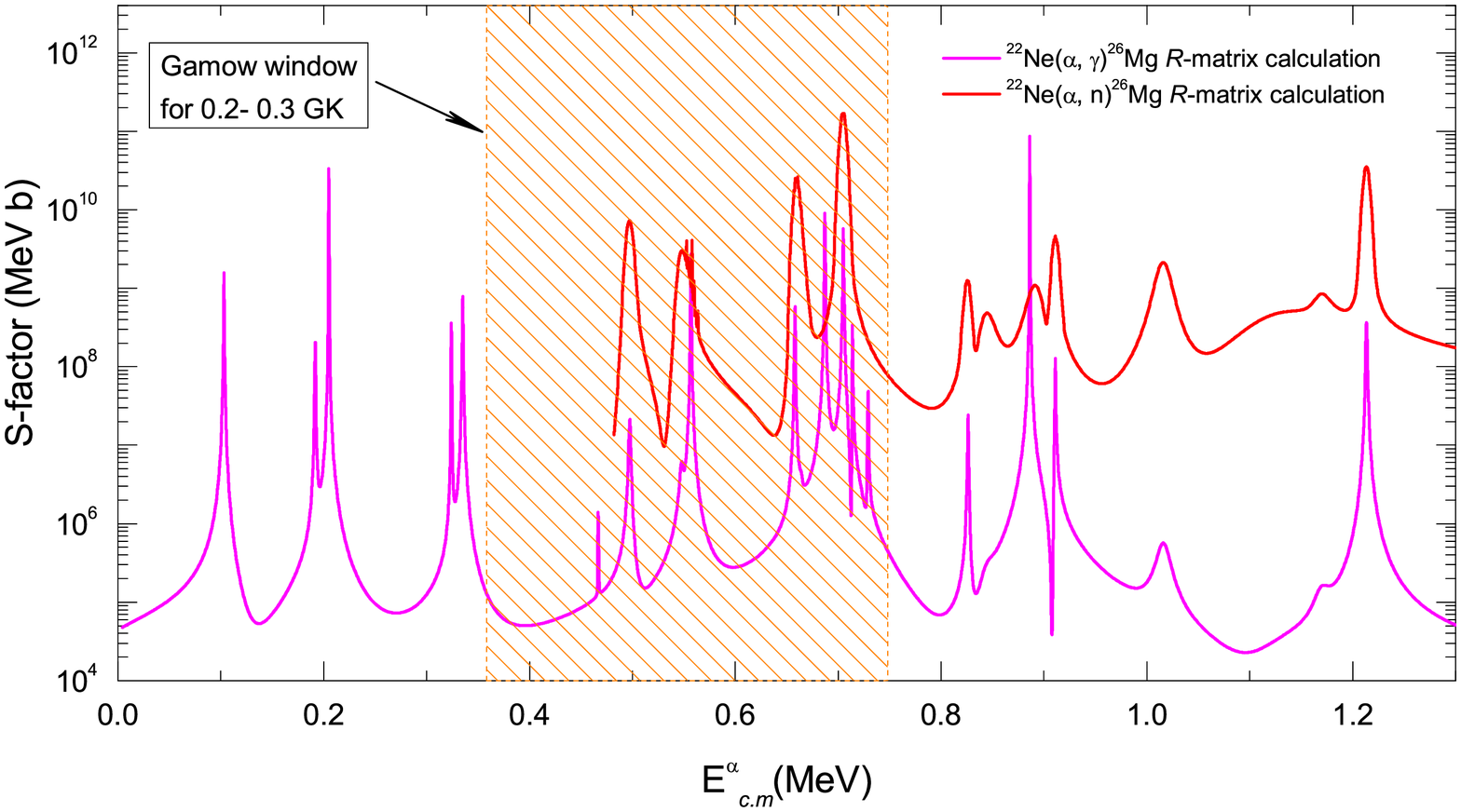}
\caption{\label{fig4} $R$-matrix calculation to the S-factor for $^{22}$Ne($\alpha$, n)$^{25}$Mg $^{22}$Ne($\alpha$, $\gamma$)$^{26}$Mg capture reaction.}
\end{center}
\end{figure}

A full R-matrix calculation has been performed for $^{22}$Ne($\alpha$, n)$^{25}$Mg with E$_{\alpha}^{c.m}$ = 0.57 to 1.45 MeV energy range and $^{22}$Ne($\alpha$, $\gamma$)$^{25}$Mg E$_{\alpha}^{c.m}$ = 0 to 1.45 MeV to evaluate reaction rate of this two reaction at Gamow window region. In this calculation E$_x$=10.69 to 11.827 MeV excitation energy all natural parity resonance states of $^{26}$Mg (0$^+$, 1$^-$, 3$^-$ 2$^+$, 4$^+$) are consider to evaluate resonance capture contribution of $^{22}$Ne($\alpha$, n)$^{25}$Mg and $^{22}$Ne($\alpha$, $\gamma$)$^{25}$Mg reaction. The resonance states of E$_x$=11.319 to 11.827 MeV (observed in measurement of $^{22}$Ne($\alpha$, n)$^{25}$Mg by Jaeger et al. \cite{Jaeger}) parameters(E$_x$, J$^\pi$, $\Gamma_{\alpha}$, $\Gamma_{n}$) are taken by $R$-matrix fitting Jaeger et al. data listed in Table \ref{tab3}. For the case of other resonance states parameter's (E$_x$, J$^\pi$, $\Gamma_{\alpha}$, $\Gamma_{n}$, $\Gamma_{\gamma}$) taken from updated nuclear data of $^{26}$Mg that are got by indirect measurement or theoretical calculation. The non resonant capture cross-section( direct capture + sub-threshold state) in $^{22}$Ne($\alpha$, $\gamma$)$^{26}$ reaction are calculated using ANC,s for different bound states of $^{26}$Mg that are extra from $^{22}$Ne($^6$Li, d)$^{26}$Mg reaction listed in Table \ref{tab1}. The states of $^{26}$Mg and their parameters used in this calculation are listed in Table \ref{tab7}. The final calculated S-factors of $^{22}$Ne($\alpha$, n)$^{25}$Mg and $^{22}$Ne($\alpha$, $\gamma$)$^{25}$Mg reaction are shown in Fig. \ref{fig4}.     

\section{Reaction rate of $^{22}$Ne($\alpha$, n)$^{25}$Mg reaction}
\begin{figure}
\begin{center}
\includegraphics[height=6.0cm, width=8.5cm]{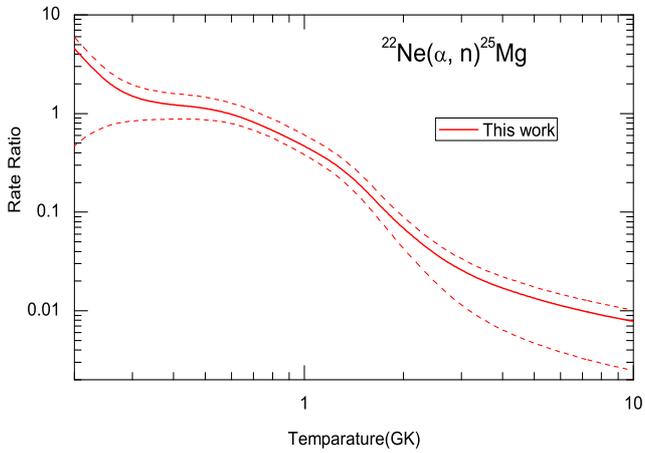}
\caption{\label{fig5} The $^{22}$Ne($\alpha$, n)$^{25}$Mg  rate of present work relative to Longland et al., meadian rate. Dash line corresponds to upper and lower limits.}
\end{center}
\end{figure}

The reaction rate of $^{22}$Ne($\alpha$, n)$^{25}$Mg reaction has been reevaluated numerically using AZURE2 code for temperature range 0.01 to 10 GK. The present rate ratio with respect to Longland et al., \cite{LONGLAND} median rate are shown in Fig. \ref{fig5} 

\section{Conclusion}
A $R$-matrix analysis in $^{22}$Ne($\alpha$, n)$^{25}$Mg reaction was performed to the description of direct measurement data \cite{Jaeger}. The present calculation with including experimental resolution corrections with recently evaluated resonance parameters \cite{Philip} 
for $^{26}$Mg are not well describe measurement data \cite{Jaeger} due to strong destructive interference between same J$^\pi$ states and as well as energy location of some resonances. With changing energy location, decay widths of some resonances and  spin, parity of E$_x$ = 11.784 and 11.63 MeV states from 1$^-$ to 0$^+$ the experimental data are well describe for E$_{\alpha}$ 0.8 to 1.45 MeV energy region. The extrapolation of $R$-matrix calculations with indirectly measured resonance parameters well off with respect to highly uncertain data of Jaeger et. al.,\cite{Jaeger} for 0.57 to 0.8 MeV range. The cross sections for $^{22}$Ne($\alpha$, $\gamma$)$^{26}$Mg are also estimated simultaneously with the resulting parameters and taking the gamma partial widths from the literature. The present $R$-matrix calculation shows that $^{22}$Ne($\alpha$, n)$^{25}$Mg is dominant with respect to $^{22}$Ne($\alpha$, $\gamma$)$^{26}$Mg reaction over the hole energy range from 0.57 to 1.2 MeV. It is also observed that the R-matrix calculation yields higher cross sections for $^{22}$Ne($\alpha$, n)$^{25}$Mg in the Gamow energy window compared to the previous estimates of Philip et al.,\cite{Philip}. The present reaction rate of $^{22}$Ne($\alpha$, n)$^{25}$Mg reaction evaluated from $R$-matrix calculation is higher by order an of magnitude as compare to Philip et al., reported at 0.2 to 0.3 GK temperature.

\end{document}